\documentclass{article}
\usepackage[dvips]{graphicx}
\begin{document}
\title{Four- and eight-fermion interactions in a space-time with non-trivial topology}
\author{Tomohiro Inagaki\\
Information Media Center, Hiroshima University, \\
1-7-1, Kagamiyama, Higashi-Hiroshima, 739-8521, JAPAN }
\date{\today}
\maketitle

\begin{abstract}
The phase structure of an eight-fermion interaction model is investigated in a 
compact space-time with non-trivial topology, $M^{D-1}\otimes S^1$.
The phase boundary dividing the symmetric and the broken phase is shown 
as a function of the space-time dimensions. It is found that the eight-fermion 
interaction does not modify the phase boundary for the chiral symmetry 
breaking. 
\end{abstract}

\section{Introduction}
It is expected that a higher symmetry may be realized at the 
early universe. Some of the symmetry  is spontaneously broken to yield a 
theory with a lower level symmetry in the current universe. The mechanism
of the symmetry breaking may be tested in critical phenomena in the early
universe. One of the possible mechanisms to induce the spontaneous
symmetry breaking is found in the dynamics of quarks and gluons. 
In QCD the expectation value of the composite operator constructed by 
a quark and an anti-quark field develops a non-vanishing value and
the chiral symmetry is eventually broken \cite{NJL}.

Y.~Nambu and G.~Jona-Lasinio introduced a four-fermion interaction
model as a prototype model to describe the dynamical mechanism of 
the symmetry breaking. The model is regards as a low energy effective
theory of a fundamental theory at high energy and extend to include 
another type interactions, higher dimensional operators and so on
\cite{tHooft,Alkofer:1990uh,HIT}.  
Much work has been done to study the phase structure of the models 
under extreme situation at the early universe, inside dense stars, or 
heavy ion collisions.

If the fundamental theory is described by the superstring theory, it is 
defined in ten space-time dimensions. Thus six space-time dimensions 
should be compactified to realize four space-time dimensions at low 
energy. The compactified space-time may have a non-trivial topology. 
The finite size and the topological effects from the compactified space-time 
may play an important role in the symmetry breaking at very early universe.
It is also expected that the critical phenomena for the symmetry breaking
may explain how to stabilize the compactification scale.

The finite size and the topological effects  are investigated in the four-fermion 
interaction model with one compactified dimension, $S^1$, 
\cite{ELO2,Vshivtsev:1995fh,IMO2}
the torus universe \cite{Kim:1987db,Song:1990dm,Kim:1994es,IIYF,Abreu:2006pt},
and D-dimensional sphere, $S^D$, \cite{IMM,IIM}.
In the case of the anti-periodic boundary condition the finite
size effect raises the ground state energy and the broken symmetry is restored 
for a sufficiently small compact direction. The finite temperature effect is 
classified as the similar type effect. The fermion
fields which possess the periodic boundary condition have an opposite
effect. The symmetry breaking is enhanced by the finite size effect.

In the present paper we consider a model with a four- and an eight-fermion 
interaction and study a contribution from the higher dimensional operator
to the finite size and the topological effects on the symmetry breaking. 
In Sec. 2 we introduce the model with scalar type four- and eight-fermion 
interactions in Minkowski space-time, $M^D$, and evaluate the effective 
potential. In Sec. 3 we consider the model in $S^1\otimes M^{D-1}$ and
discuss the influence of the finite-size and the non-trivial topology to the
symmetry breaking. Solving the gap equations, we calculate the critical size
of the compactified dimension. 
The critical temperature is also mentioned.
In Sec. 4 we give concluding remarks.

\section{Eight-fermion interaction model in Minkowski space-time, $M^{D}$}
In this section we introduce the model with scalar type four- and eight-fermion 
interactions and study the features of the model in arbitrary space-time dimensions
$2<D<4$. We start with the simple model with $N$ components of fermions 
described by the Lagrangian \cite{HIT},
\begin{equation}
  {\cal L}=\sum_{i=1}^{N}
    \bar{\psi}_i i\gamma^\mu\partial_\mu\psi_i
    + \frac{G_1}{N} \left(\sum_{i=1}^{N} \bar{\psi}_i\psi_i\right)^2
    + \frac{G_2}{N} \left(\sum_{i=1}^{N} \bar{\psi}_i\psi_i\right)^4,
\label{lag:org}
\end{equation}
where the index $i$ represents flavor of the fermion field $\psi$,
$N$ is the number of the flavors, $G_1$ and $G_2$ the coupling 
constants for the four- and the eight-fermion interactions, 
respectively.

The Lagrangian (\ref{lag:org}) has a global $SU(N)$ flavor symmetry under
the transformation,
\begin{equation}
  \psi\rightarrow e^{i\sum_a g_a T^a}\psi,
\end{equation}
where $T^a$ is the generators of the $SU(N)$ symmetry.
The Lagrangian (\ref{lag:org}) is also invariant under a discrete $Z_2$ 
chiral transformation,
\begin{equation}
  \psi\rightarrow \gamma_5\psi.
\end{equation}
This discrete chiral symmetry prevents the Lagrangian from having mass terms.
If the composite operator $\bar{\psi}\psi$ develops a non-vanishing value, 
fermion mass terms are generated and the discrete chiral symmetry is 
dynamically broken.

In practical calculations it is more convenient to introduce auxiliary
fields, $\sigma_1$, $\sigma_2$, and evaluate the equivalent Lagrangian
\begin{equation}
{\cal L}_{aux}=
\bar{\psi}\left(i\gamma^\mu\partial_\mu -\sigma\right)\psi 
-\frac{N \sigma_1^2}{4G_1}-\frac{N \sigma_2^2}{4G_2}
\label{lag:aux},
\end{equation}
where $\sigma$ is defined by
\begin{equation}
  \sigma\equiv
  \sigma_1 \sqrt{1+\left|\frac{N\sigma_2}{G_1}\right|}.
\end{equation}
It is easily found that the original Lagrangian (\ref{lag:org}) can 
be reproduced by replacing $\sigma_1$ and $\sigma_2$ by the solutions of the
equations of motion,
\begin{equation}
  \sigma_1 = -\frac{2G_1}{N}\sqrt{1+\left|\frac{N\sigma_2}{G_1}\right|}\bar{\psi}\psi ,\,\,\,
  |\sigma_2| = \mbox{sgn}(G_1)\frac{2G_2}{N}(\bar{\psi}\psi)^2 .
\label{eom}
\end{equation}

To find the ground state of the model we observe the minimum of the effective
potential for the auxiliary fields $\sigma_1$ and $\sigma_2$. In the leading order 
of the $1/N$ expansion the effective potential is given by
\begin{equation}
  V_0(\sigma_1,\sigma_2)=
    \frac{\sigma_1^2}{4G_1}+\frac{\sigma_2^2}{4G_2}
    +i\mbox{tr}\ln
    \frac{i\gamma^\mu\partial_\mu-\sigma}{i\gamma^\mu\partial_\mu},
\label{v0:def}
\end{equation}
where we normalize the effective potential so that $V(0,0)=0$.

In the $D$-dimensional Minkowski space-time, $M^D$, we rewrite the second
term on the right hand side of Eq.(\ref{v0:def}) by using the Schwinger
proper time method and obtain
\begin{equation}
  V_0=
    \frac{\sigma_1^2}{4G_1}+\frac{\sigma_2^2}{4G_2}
    -i\mbox{tr}\int_0^\sigma sds\int
    \frac{d^{D}k}{(2\pi)^{D}}\frac{1}{k^2-s^2}.
\label{v0:org}
\end{equation}
We can perform the integration and get 
\begin{eqnarray}
  V_0&=&
    \frac{\sigma_1^2}{4G_1}+\frac{\sigma_2^2}{4G_2}
    +\frac{\mbox{tr}1}{2(4\pi)^{D/2}}\Gamma\left(-\frac{D}{2}\right)(\sigma^2)^{D/2}
\nonumber\\
    &=&\frac{\sigma_1^2}{4G_1}+\frac{\sigma_2^2}{4G_2}
    +\frac{\mbox{tr}1}{2(4\pi)^{D/2}}\Gamma\left(-\frac{D}{2}\right)({\sigma_1}^2)^{D/2}
    \left( 1+\left|\frac{N\sigma_2}{G_1}\right| \right)^{D/2}.
\label{v0:int}
\end{eqnarray}
Since the first and the third terms on the right hand side of Eq.(\ref{v0:int}) disappear for 
$\sigma_1=0$, 
we should set a positive value for the coupling constant $G_2$ to obtain a stable ground 
state. At the limit $\sigma_2\rightarrow 0$ the effective potential (\ref{v0:int}) coincides with 
the one for the four-fermion interaction model.
If the space-time dimensions are more than four, the main contribution comes from the third 
term on the right hand side of Eq.(\ref{v0:int}) for a large $\sigma_1$ and $\sigma_2$. 
Thus the effective potential can not be bounded below and there is no stable ground state. 
For $6<D<8$ a stable ground state can be realized.

To find the minimum of the effective potential, $V_0$, we differentiate the effective
potential in terms of $\sigma_1$ and $\sigma_2$ and get
\begin{equation}
  \frac{\partial V_0}{\partial \sigma_1}=\sigma_1\left[
    \frac{1}{2G_1}
    -\frac{\mbox{tr}1}{(4\pi)^{D/2}}\Gamma\left(1-\frac{D}{2}\right)({\sigma_1}^2)^{D/2-1}
    \left( 1+\left|\frac{N\sigma_2}{G_1}\right| \right)^{D/2}\right],
\label{gap0:1}
\end{equation}
\begin{equation}
  \frac{\partial V_0}{\partial \sigma_2}=
    \frac{\sigma_2}{2G_2}
    -\frac{\mbox{sgn}({\sigma_2})\mbox{tr}1}{2(4\pi)^{D/2}}\Gamma\left(1-\frac{D}{2}\right)
    \frac{N}{|G_1|}({\sigma_1}^2)^{D/2} \left( 1+\left|\frac{N\sigma_2}{G_1}\right| \right)^{D/2-1}.
\label{gap0:2}
\end{equation}
For $2<D<4$ and $6<D<8$ the right hand side of Eq.(\ref{gap0:2}) has a negative value 
for $\sigma_2<0$ and a positive value for $\sigma_2>0$. It implies that the effective potential 
monotonically decreases for $\sigma_2<0$ and increases for $\sigma_2>0$.
Therefore the minimum of the effective potential is found at $\sigma_2=0$ and the 
expectation value for $\sigma_1$ is equal to that for the four-fermion interaction model,
\begin{equation}
  \langle\sigma_1\rangle=\pm\left[
    \frac{1}{2G_1}\frac{(4\pi)^{D/2}}{\mbox{tr}1\displaystyle\Gamma\left(1-D/2\right)}
  \right]^{1/(D-2)}.
\end{equation}

\begin{figure}[htbp]
 \begin{minipage}{0.49\hsize}
  \begin{center}
   \includegraphics[width=60mm]{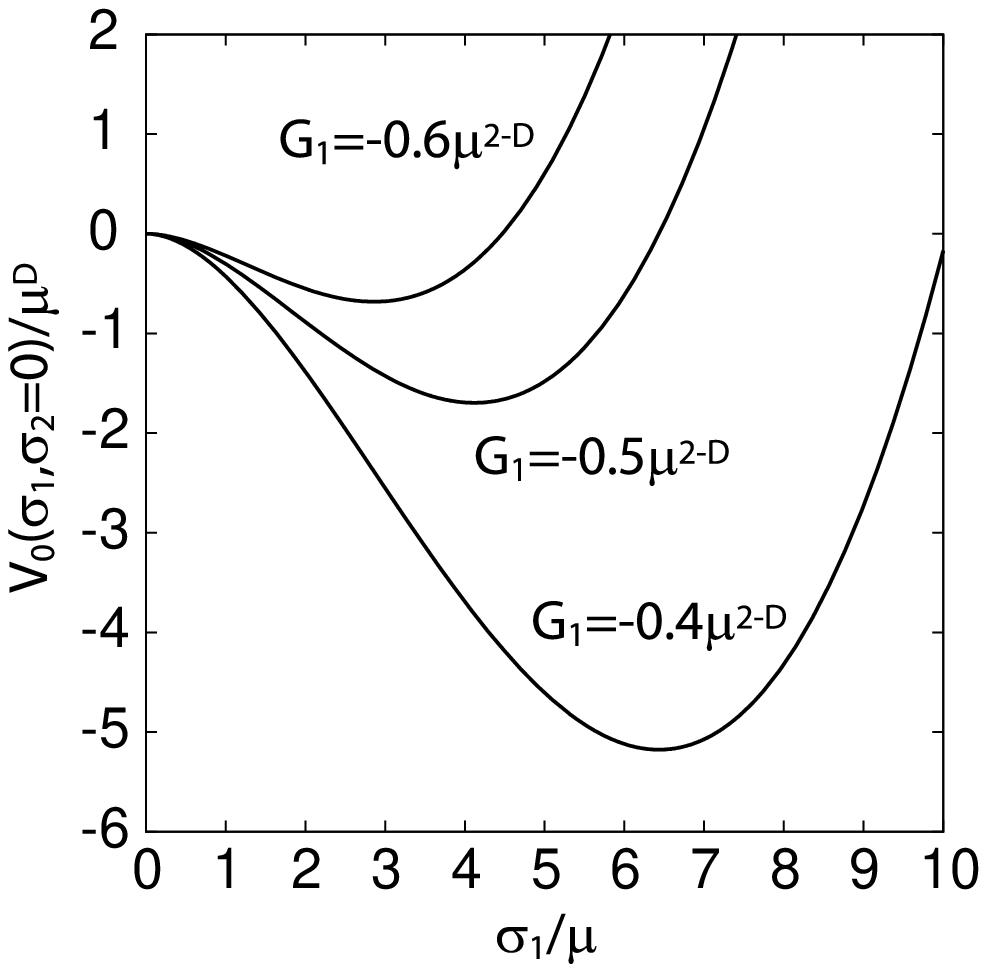}
   \caption{{\footnotesize Behavior of the effective potential (\ref{v0:int}) 
  at $\sigma_2=0$ for $N=3$, $G_2=0.5\mu^{4-3D}$, $D=2.5$.}}
     \label{v01}
     \end{center}
  \end{minipage}
  \begin{minipage}{0.02\hsize}
  \end{minipage}
  \begin{minipage}{0.49\hsize}
  \begin{center}
   \includegraphics[width=60mm]{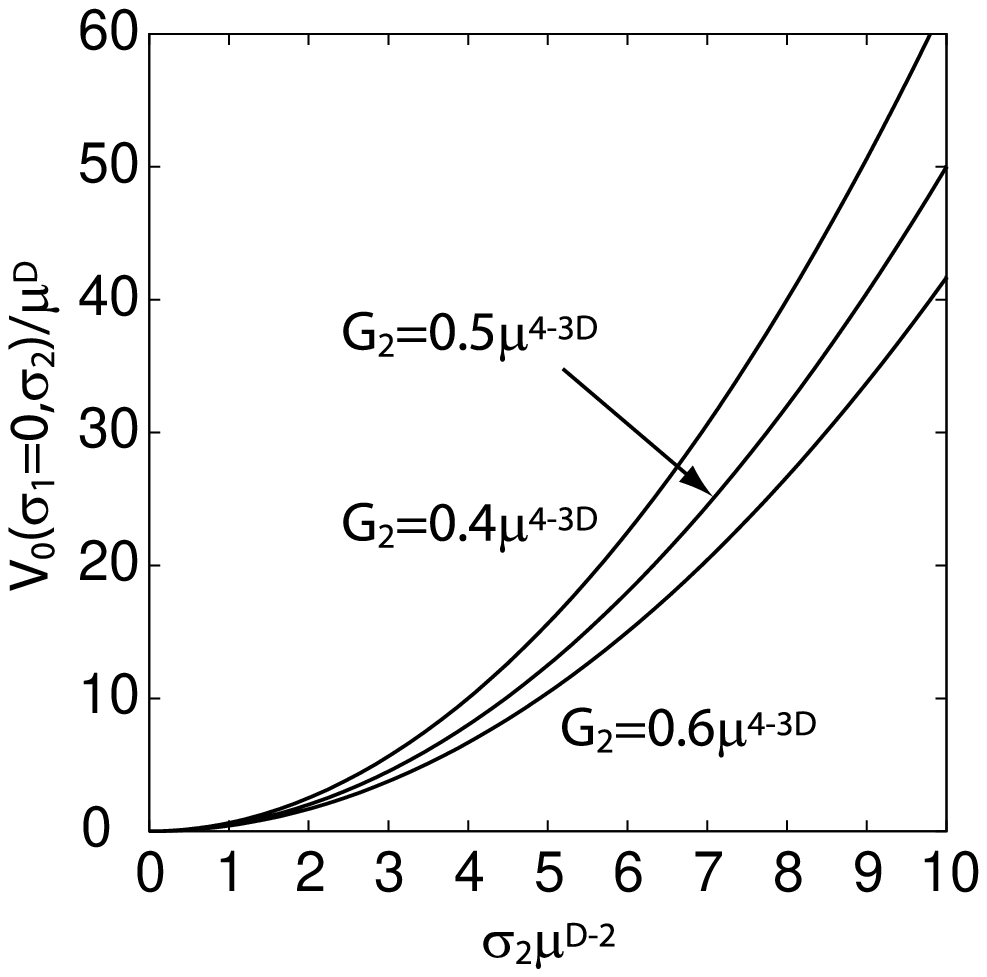}
   \caption{{\footnotesize Behavior of the effective potential (\ref{v0:int})
   at $\sigma_1=0$ for $N=3$, $G_1=-0.5\mu^{2-D}$, $D=2.5$.}}
   \label{v02}
   \end{center}
 \end{minipage}
\end{figure}

We normalize the mass scale by an arbitrary renormalization scale $\mu$ and 
numerically evaluate the effective potential (\ref{v0:int}).
We observe only the symmetric phase for a positive $G_1$.
The discrete chiral symmetry is broken for $G_1<0$.
In Figs.\ref{v01} and \ref{v02} a typical behavior of the effective potential, 
$V_0(\sigma_1, \sigma_2)$, is drawn for fixed space-time dimension $D$ 
less than four.
As is shown in Fig.\ref{v01}, the state $\sigma_1=\sigma_2=0$ is unstable
for a negative four-fermion coupling $G_1$. Thus the composite operator,
$\bar{\psi}\psi$ develops a non-vanishing value and the discrete chiral 
symmetry is dynamically broken. This result exactly reproduce the one obtained
in Ref.\cite{IKM}.
In Fig.\ref{v02}  it is shown that the eight-fermion interaction enhances the 
symmetry breaking. We also observe that the expectation value for the auxiliary 
field, $\sigma_2$, is vanishing. It is consistent with the behavior of Eq.(\ref{gap0:2}) 
discussed above.

\section{Eight-fermion interaction model in $M^{D-1}\otimes S^1$}

If the space-time has a compact direction, the effective potential is modified 
by finite size and topological effect. Here we focus on the finite size and 
topological effect and consider a simple flat space-time with non-trivial 
topology. We assume that one of the space direction is one dimensional 
sphere, $S^1$. In $M^{D-1}\otimes S^1$ the effective potential is given by
\begin{equation}
  V(\sigma_1,\sigma_2)=
    \frac{\sigma_1^2}{4G_1}+\frac{\sigma_2^2}{4G_2}
    -i\mbox{tr}\int_0^\sigma sds\frac{1}{L}\sum_{n=-\infty}^{\infty}\int
    \frac{d^{D-1}k}{(2\pi)^{D-1}}\frac{1}{k^2-s^2},
\label{v:org}
\end{equation}
where $L$ is the size of the $S^1$ direction, $k_1$ is replaced as
\begin{equation}
  k_1 \rightarrow \frac{(2n+\delta_{p,1})\pi}{L},
\end{equation}
and $\delta_{p,1}$ is fixed by the boundary condition, i.e. $\delta_{p,1}=0$
for the fermion fields with the periodic boundary condition and $\delta_{p,1}=1$
for the fermion fields with the anti-periodic boundary condition \cite{ELO2, IMO2}. 

Performing the integration in Eq.(\ref{v:org}), the effective potential reads
\begin{eqnarray}
  V(\sigma_1,\sigma_2)&=&
    \frac{\sigma_1^2}{4G_1}+\frac{\sigma_2^2}{4G_2}
    +\frac{\mbox{tr}1}{2(4\pi)^{(D-1)/2}}\Gamma\left(\frac{1-D}{2}\right)
\nonumber \\
    &&\times \frac{1}{L}\sum_{n=-\infty}^{\infty}\left[
    \left(\frac{(2n+\delta_{p,1})\pi}{L}\right)^2+\sigma^2\right]^{(D-1)/2}
    -V(0,0).
\label{v:int}
\end{eqnarray}
This expression is convenient for analytically evaluating the behavior of the
effective potential. 
Only the second term on the right hand side of Eq.(\ref{v:int}), 
remains for $\sigma_1=0$, we again set a positive value for the coupling 
constant $G_2$ to obtain a stable ground state. 

We can also perform the summation in Eq.(\ref{v:org}) and get
\begin{eqnarray}
  V(\sigma_1,\sigma_2)&=&V_0(\sigma_1,\sigma_2)
    -\frac{\mbox{tr}1}{(4\pi)^{(D-1)/2}}\frac{2}{\displaystyle
    \Gamma\left(\frac{D-1}{2}\right)}
\nonumber \\
    &&\times \frac{1}{L}\int_0^{\infty} k^{D-2}dk\ln\frac{1-(-1)^{\delta_{p,1}}
    \exp\left(-L\sqrt{k^2+\sigma^2}\right)}
    {1-(-1)^{\delta_{p,1}}\exp\left(-L|k|\right)}.
\label{v:int:perant}
\end{eqnarray}
These expressions is suitable for the numerical analysis of the effective potential.

To find the phase structure of the model we want to find the ground state of the
model in $S^1\otimes M^D$.
The ground state is found by observing the minimum of the effective potential.
The potential is not always analytic at $\sigma_2=0$. Hence, the minimum 
satisfies the equations,
\begin{eqnarray}
  \frac{\partial V}{\partial \sigma_1} &=&
    \frac{\sigma_1}{2G_1}
    -\frac{\mbox{tr}1}{(4\pi)^{(D-1)/2}}\Gamma\left(\frac{3-D}{2}\right)
    \sigma_1\left(1+\left|\frac{N\sigma_2}{G_1}\right|\right)
\nonumber \\
    &&\times \frac{1}{L}\sum_{n=-\infty}^{\infty}\left[
    \left(\frac{(2n+\delta_{p,1})\pi}{L}\right)^2+\sigma^2\right]^{(D-3)/2}
=0,
\label{gap:1}
\end{eqnarray}
and
\begin{eqnarray}
  \frac{\partial V}{\partial \sigma_2} &=&
    \frac{\sigma_2}{2G_2}
    -\frac{\mbox{sgn}(\sigma_2)\mbox{tr}1}{2(4\pi)^{(D-1)/2}}\Gamma\left(\frac{3-D}{2}\right)
    \frac{N{\sigma_1}^2}{|G_1|}
\nonumber \\
    &&\times \frac{1}{L}\sum_{n=-\infty}^{\infty}\left[
    \left(\frac{(2n+\delta_{p,1})\pi}{L}\right)^2+\sigma^2\right]^{(D-3)/2}
=0,
\label{gap:2}
\end{eqnarray}
or $\sigma_2=0$. In the latter case, $\sigma_2=0$, the effective potential
coincides with the one for the four-fermion interaction model and the minimum
of the effective potential is found by solving Eq.(\ref{gap:1}) at $\sigma_2=0$.

Non-vanishing expectation values for $\sigma_1$ and $\sigma_2$ are given by
the solution of the gap equations, 
\begin{eqnarray}
    \frac{1}{2G_1}&=&
    \frac{\mbox{tr}1}{(4\pi)^{(D-1)/2}}\Gamma\left(\frac{3-D}{2}\right)
    \left(1+\left|\frac{N\langle\sigma_2\rangle}{G_1}\right|\right)
\nonumber \\
    &&\times \frac{1}{L}\sum_{n=-\infty}^{\infty}\left[
    \left(\frac{(2n+\delta_{p,1})\pi}{L}\right)^2+
    \langle\sigma\rangle^2\right]^{(D-3)/2},
\label{gap:non1}
\end{eqnarray}
and
\begin{eqnarray}
    \frac{1}{2G_2}&=&
    \frac{\mbox{tr}1}{2(4\pi)^{(D-1)/2}}\Gamma\left(\frac{3-D}{2}\right)
    \frac{N{\langle\sigma_1\rangle}^2}{|G_1\langle\sigma_2\rangle|}
\nonumber \\
    &&\times \frac{1}{L}\sum_{n=-\infty}^{\infty}\left[
    \left(\frac{(2n+\delta_{p,1})\pi}{L}\right)^2
    +\langle\sigma\rangle^2\right]^{(D-3)/2}.
\label{gap:non2}
\end{eqnarray}
Eliminating the summation from these gap equations, we obtain a 
relationship between $\langle \sigma_1\rangle$ and $\langle \sigma_2\rangle$ 
\cite{HIT},
\begin{equation}
  \frac{N}{2G_1|G_1|}\langle\sigma_1\rangle^2=
  \frac{|\langle\sigma_2\rangle|}{G_2}
  \left(1+N\left|\frac{\langle\sigma_2\rangle}{G_1}\right|\right).
\label{rel:s1s2}
\end{equation}

\begin{figure}[htbp]
 \begin{minipage}{0.49\hsize}
  \begin{center}
  \includegraphics[width=60mm]{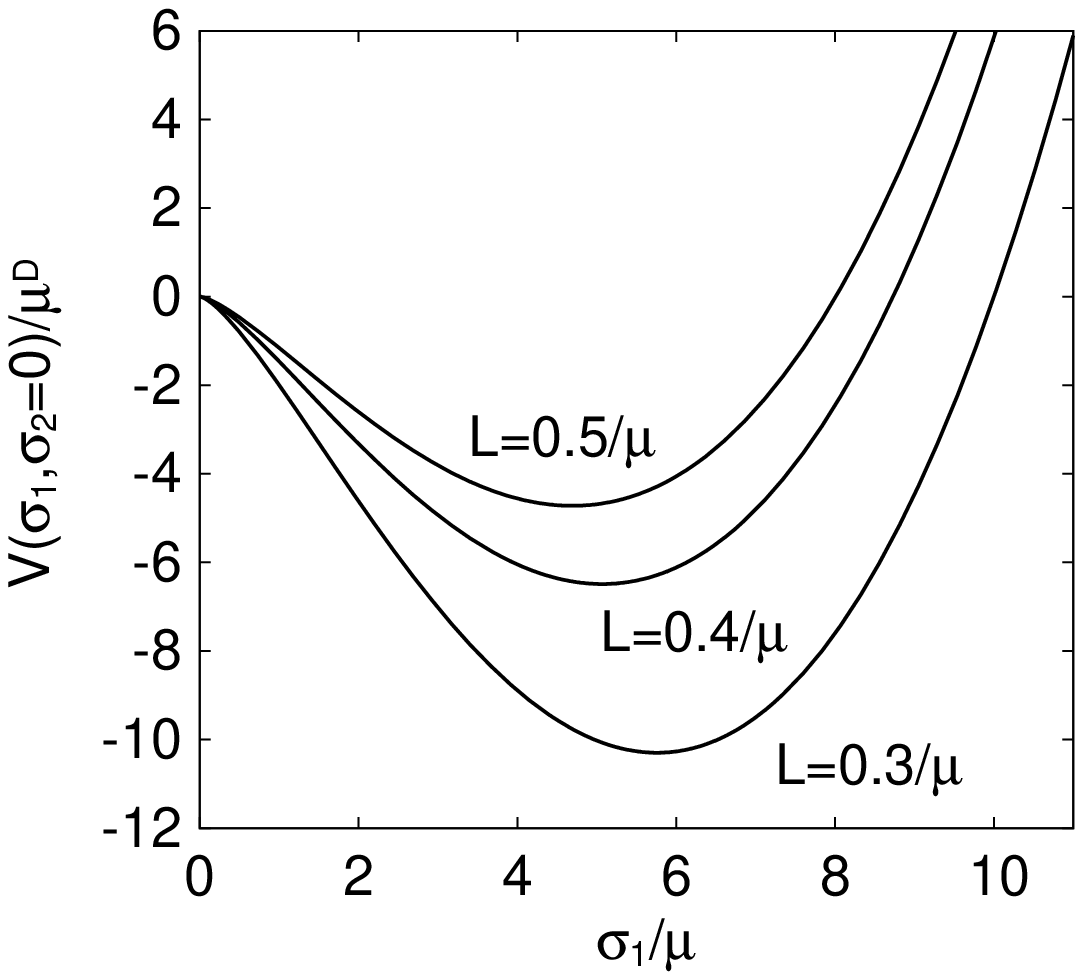}
  \caption{{\footnotesize Behavior of the effective potential (\ref{v:int:perant})
   for fermion fields with the periodic boundary condition, $\delta_{p,1}=0$,
   at $\sigma_2=0$ for $N=3$, $G_1=-0.5\mu^{2-D}$, $G_2=0.5\mu^{4-3D}$, $D=2.5$.}}
  \label{v1}
  \end{center}
  \end{minipage}
  \begin{minipage}{0.02\hsize}
  \end{minipage}
  \begin{minipage}{0.49\hsize}
  \begin{center}
  \includegraphics[width=60mm]{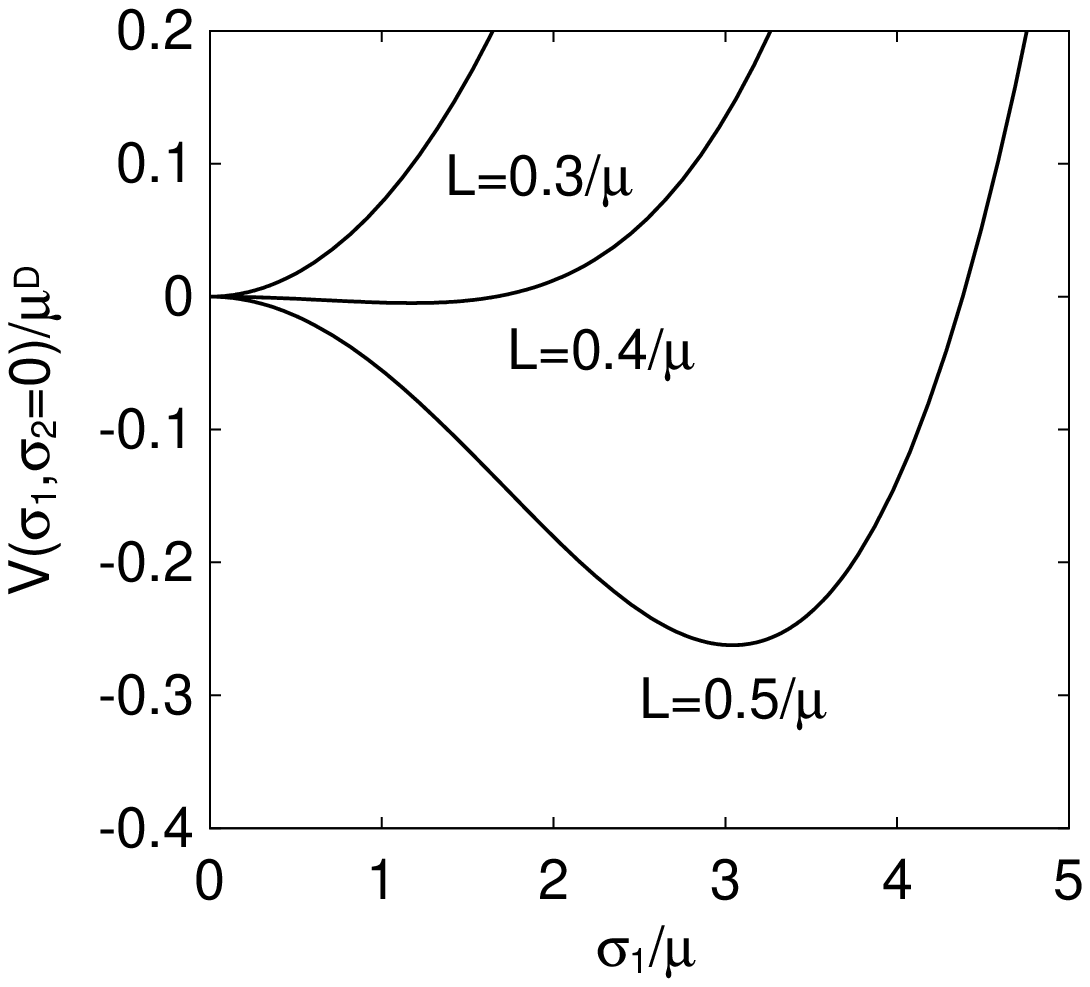}
  \caption{{\footnotesize Behavior of the effective potential (\ref{v:int:perant}) 
  for fermion fields with the anti-periodic boundary condition, $\delta_{p,1}=1$,
  at $\sigma_2=0$ for $N=3$, $G_1=-0.5\mu^{2-D}$, $G_2=0.5\mu^{4-3D}$, $D=2.5$.}}
 \label{v2}
 \end{center}
 \end{minipage}
\end{figure}

\begin{figure}[htbp]
 \begin{minipage}{0.49\hsize}
  \begin{center}
   \includegraphics[width=60mm]{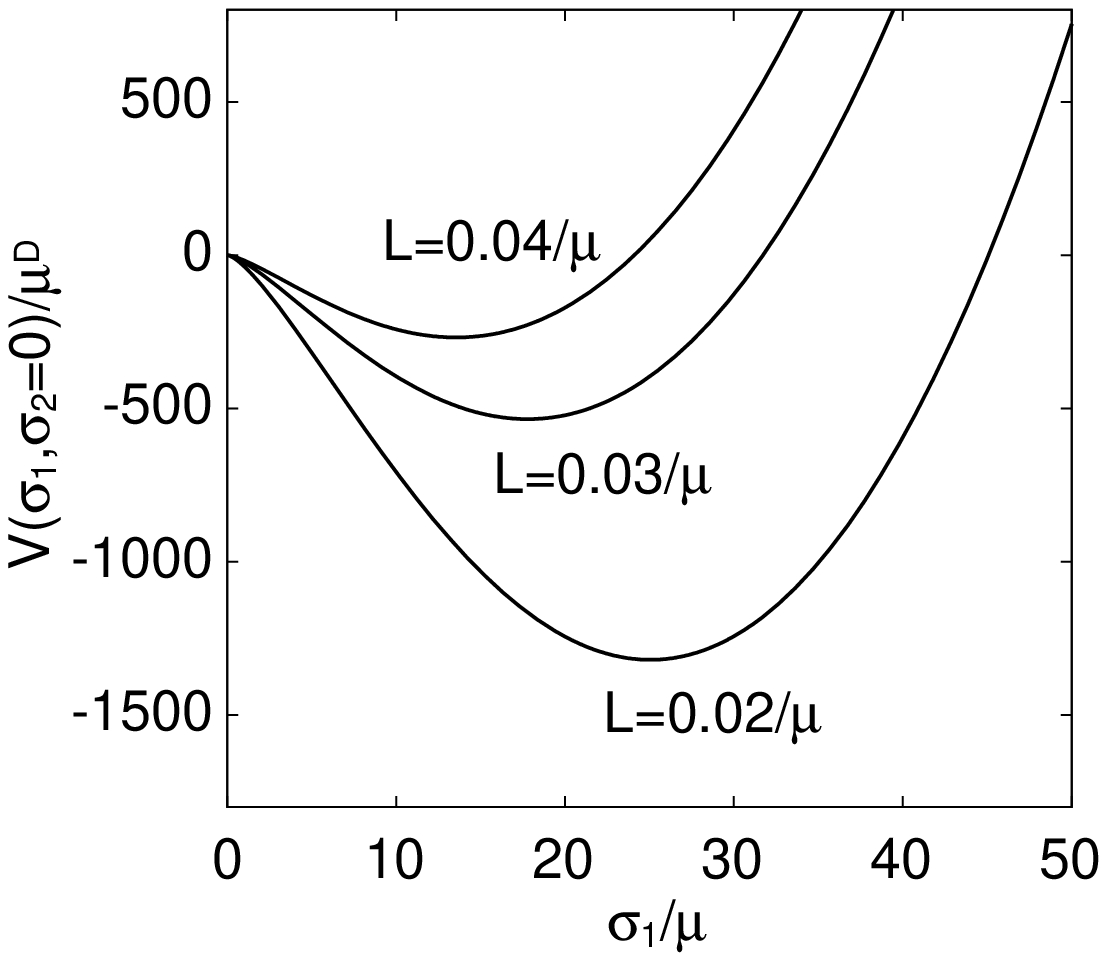}
   \caption{{\footnotesize Behavior of the effective potential (\ref{v:int:perant}) 
  for fermion fields with the periodic boundary condition, $\delta_{p,1}=0$,
  at $\sigma_2=0$, $N=3$, $G_1=0.5\mu^{2-D}$, $G_2=0.5\mu^{4-3D}$, $D=2.5$.}}
     \label{v3}
     \end{center}
  \end{minipage}
  \begin{minipage}{0.02\hsize}
  \end{minipage}
  \begin{minipage}{0.49\hsize}
  \begin{center}
   \includegraphics[width=60mm]{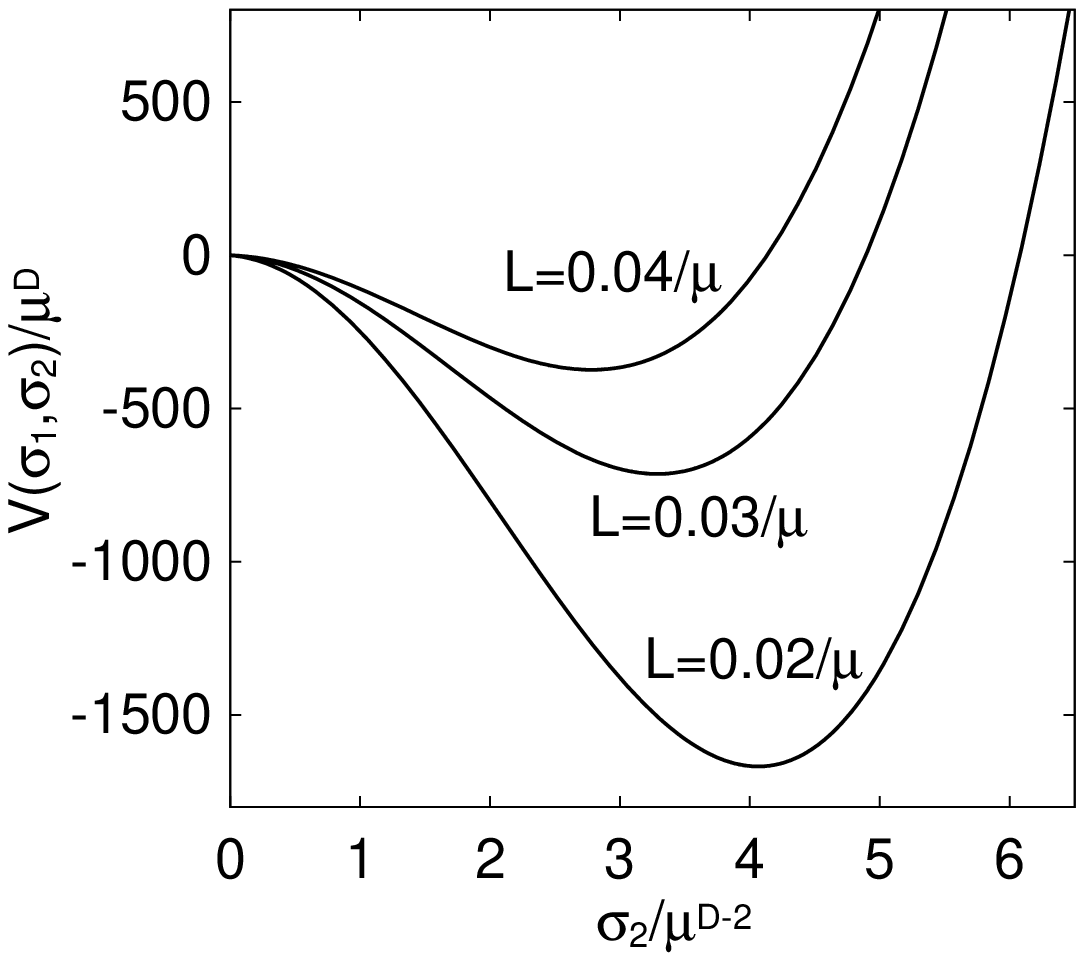}
   \caption{{\footnotesize Behavior of the effective potential (\ref{v:int:perant})
   for fermion fields with the periodic boundary condition , $\delta_{p,1}=0$,
   under the relationship (\ref{rel:s1s2})
   at $N=3$, $G_1=0.5\mu^{2-D}$, $G_2=0.5\mu^{4-3D}$, $D=2.5$.}}
   \label{v4}
   \end{center}
 \end{minipage}
\end{figure}

If we set a negative value to  $G_1$, Eq.(\ref{rel:s1s2}) has no solution and the 
minimum is given by Eq.(\ref{gap:non2}) at $\sigma_2=0$. 
We numerically show the effective potential at $\sigma_2=0$ as a
function of $\sigma_1$ in Figs.\ref{v1} and \ref{v2} for fermion fields 
with the periodic and the anti-periodic boundary condition, respectively.
As is shown in Fig.\ref{v1}, the symmetry breaking is 
enhanced by the finite size effect and only the broken phase can be 
realized for the fermion fields with the periodic boundary condition.
In Fig.\ref{v2} it is clearly seen that the finite size effect restores the broken 
symmetry through the second order phase transition for the fermion fields 
with the anti-periodic boundary condition. 
These results exactly reproduce those for the four-fermion interaction model.
 
We also numerically evaluate the effective potential for a positive $G_1$.
In this case we observe only the symmetric phase for the fermion fields 
with the anti-periodic boundary condition. On the other hand, we find
the broken phase for the fermion fields with the periodic boundary 
condition. The chiral symmetry is dynamically broken for a sufficiently 
small $L$ through the second order phase transition.
In Fig.\ref{v3} we draw the effective potential at $\sigma_2=0$. We also draw the
effective potential under the relationship (\ref{rel:s1s2}) in Fig.\ref{v4}. Comparing
Fig.\ref{v3} with \ref{v4}, we find that the absolute minimum of the effective potential 
satisfies the relationship (\ref{rel:s1s2}). It implies that the eight-fermion interaction
enhances the symmetry breaking. 

Since the phase transition is of the second order, the critical length, $L_{cr}$, 
is obtained by taking the limit  $\langle\sigma_1\rangle\rightarrow 0$ and 
$\langle\sigma_2\rangle\rightarrow 0$ in Eqs. (\ref{gap:non1}) and 
(\ref{gap:non2}). We can perform the summation in these equations at the 
limit and get
\begin{equation}
  L_{cr}=
  2\pi\left[\frac{2\mbox{tr}1G_1}{\pi(4\pi)^{(D-1)/2}}
  \Gamma\left(\frac{3-D}{2}\right)\zeta(3-D)\right]^{1/(D-2)},
\label{lcr:1}
\end{equation}
for fermion fields with the periodic boundary condition and
\begin{equation}
  L_{cr}=
  2\pi\left[\frac{2\mbox{tr}1G_1}{\pi(4\pi)^{(D-1)/2}}(2^{3-D}-1)
  \Gamma\left(\frac{3-D}{2}\right)\zeta(3-D)\right]^{1/(D-2)},
\label{lcr:2}
\end{equation}
for fermion fields with the anti-periodic boundary condition.
Since the critical length, $L_{cr}$, depends only on 
the four-fermion coupling $G_1$ and the space-time dimensions $D$, 
the eight-fermion coupling has no contribution to the phase boundary.
The critical length coincides with the four-fermion interaction model \cite{IMO2}.
The model for fermion fields with the anti-periodic boundary condition in 
$S^{1}\otimes M^{D-1}$ is equivalent to the model at finite temperature, $T=1/L$.
Thus the critical temperature of our model is given by the inverse of Eq.(\ref{lcr:2}).

\begin{figure}[htbp]
 \begin{minipage}{0.49\hsize}
  \begin{center}
   \includegraphics[width=60mm]{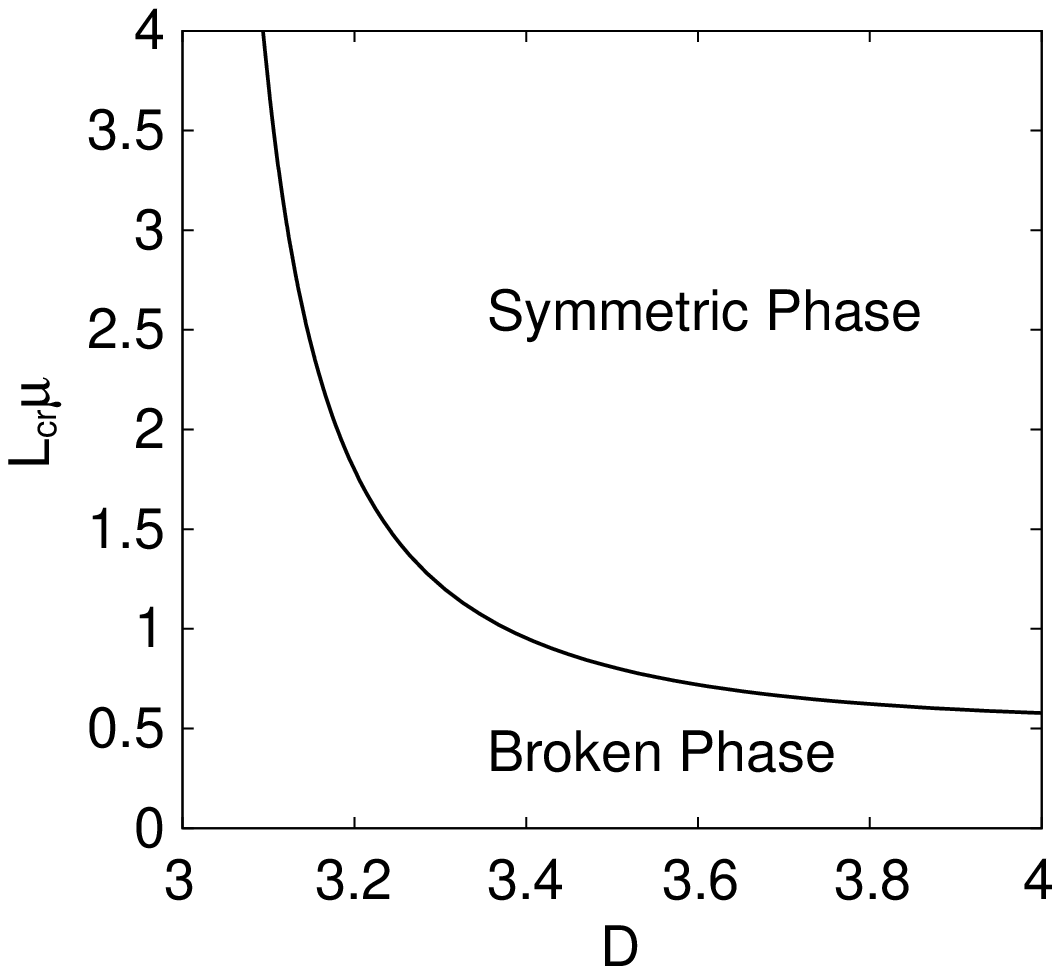}
   \caption{{\footnotesize Phase diagram for fermion fields with the periodic boundary condition
   at $G_1=0.5\mu^{2-D}$}}
   \label{cL:Pe}
   \end{center}
  \end{minipage}
  \begin{minipage}{0.02\hsize}
  \end{minipage}
  \begin{minipage}{0.49\hsize}
  \begin{center}
   \includegraphics[width=60mm]{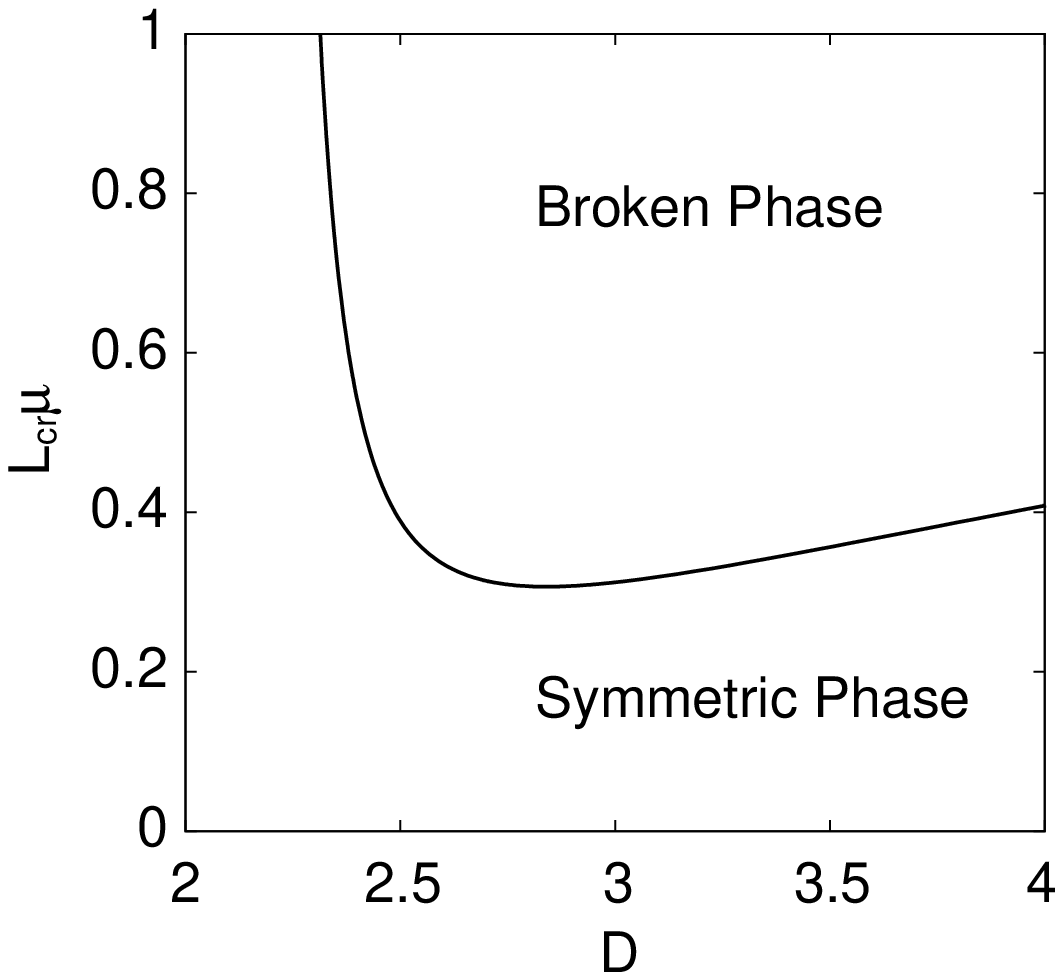}
   \caption{{\footnotesize Phase diagram for fermion fields with the anti-periodic boundary condition
   at $G_1=-0.5\mu^{2-D}$}}
   \label{cL:An}
   \end{center}
 \end{minipage}
\end{figure}

We plot the critical length of the compactified dimension for a (\ref{lcr:1}) and
(\ref{lcr:2}), as a function of the space-time dimensions in Figs.\ref{cL:Pe} and 
\ref{cL:An}, respectively. Blow three dimensions Eq.(\ref{lcr:2}) has no real
solution for a positive $G_1$. Thus the critical length is divergent at the limit,
$D\rightarrow 3$, in Fig.\ref{cL:Pe}. Because of the topological effect from the
fermion fields with the periodic boundary condition only the broken phase can
be realized for $2<D<3$. To find a finite critical length in Fig.\ref{cL:An} at 
the limit $D\rightarrow 2$ we have to renormalize the four-fermion coupling, $G_1$.
The renormalization of the coupling is discussed in Ref.\cite{IKM}.

\section{Conclusion}

We have investigated the model with the scalar type four- and eight fermion interactions
in Minkowski space-time, $M^D$, and a space-time with non-trivial topology, 
$S^1\otimes R^{D-1}$. Evaluating the effective potential, we have shown the finite size 
and topological effects on the phase structure of the chiral symmetry breaking.

In D-dimensional Minkowski space-time, $M^D (2<D<4)$, the minimum of the effective 
potential is equivalent to the one for the four-fermion interaction model.
The discrete chiral symmetry is dynamically broken for a negative four-fermion
coupling, $G_1$.
For $2<D<4$ the four-fermion interaction is marginal but the eight-fermion
interaction is irrelevant. Since the irrelevant operator has only a small contribution
at low energy, the phase structure for the model coincides with the one for the
four-fermion interaction model in $M^D$.

In $S^1\otimes R^{D-1}$ the opposite finite size effects are observed between 
the fermion fields with the periodic and the anti-periodic boundary conditions. 
Under the periodic boundary condition the symmetry breaking is enhanced
and only the broken phase can be realize for a negative $G_1$. It is found that 
the discrete chiral symmetry is dynamically broken through the second order
phase transition by the finite size effect for a positive $G_1$. In this case it is 
observed that the eight-fermion interaction also enhances the symmetry breaking.
In the case of the anti-periodic boundary condition the symmetry breaking
is suppressed. Only the symmetric phase can be realized for a positive $G_1$.
For a negative $G_1$ the broken symmetry is restored as the length of the
compactified space decreases. The phase transition is of the second order.

It should be noted the regularization dependence of the result. We can regard the 
model in $D$ dimensions as the regularized model of the low energy
effective theory in four-dimensions and compare the result with the one with cut-off 
regularization \cite{HIT}. Since the four- and eight- fermion interactions are 
non-renoramalizable in four dimensions, some of physical behaviors depend on the
regularization procedure. The result obtained in the present paper may have difference
from the one obtained in the other regularization procedure.
In the cut-off regularized model, the eight-fermion interaction has non-negligible effect
on the phase structure even in $M^4$.

One is interested in applying the result to the model with extra dimensions.
If the space-time dimensions are no less than four, both the four- and eight-fermion
interactions are irrelevant. It is expected that the eight-fermion interaction has 
non-negligible contributions. 
It would be interesting also to study such effects in supersymmetric models in 
curved space-time \cite{BIO}. One will consider the problem further and hope
to publish reports in these directions.

\section*{Acknowledgements}
The author is supported by the Ministry of Education, Science, Sports and 
Culture, Grant-in-Aid for Scientific Research (C), No. 18540276, 2008.

\end{document}